\documentclass[12pt]{iopart}

\usepackage{graphicx}
\usepackage{color}

\newcommand{\bma}[1]{\mbox{\boldmath${#1}\/$}}
\newcommand{\Cdot}{\bma{\cdot}}
\newcommand{\Nabla}{\bma{\nabla}}

\begin{document}
\title{Interpolation of the magnetic field at the test masses in eLISA}

\author{I Mateos$^{1,2}$, 
        M D\'\i az--Aguil\'o$^1$, 
        J Ramos--Castro$^{1,3}$,\\ 
        E\ Garc\'\i a--Berro$^{1,4}$ and 
        A Lobo$^{1,2}$\footnote{Deceased}}

\address{$^{1}$Institut d'Estudis Espacials de Catalunya, 
               c/Gran Capit\`a 2--4, 
               Edif. Nexus 201, 
               08034 Barcelona,
               Spain}
\address{$^{2}$Institut de Ci\`encies de l'Espai (CSIC), 
               Campus de la UAB, 
               Facultat de Ci\`encies, 
               Torre C5 -- parell, 
               08193 Bellaterra, 
               Spain}
\address{$^{3}$Departament d'Enginyeria Electr\`onica,  
               Universitat Polit\`ecnica de Catalunya, 
               c/Jordi Girona 1--3, 
               Edifici C4,
               08034 Barcelona, 
               Spain}
\address{$^{4}$Departament de F\'isica Aplicada, 
               Universitat Polit\`ecnica de Catalunya, 
               c/Esteve Terrades 5, 
               08860 Castelldefels, 
               Spain}

\ead{mateos@ice.csic.es}

\begin{abstract}
A feasible design for a magnetic diagnostics subsystem for eLISA  will be based on that of
its precursor mission, LISA  Pathfinder. Previous experience indicates
that magnetic field estimation at the positions of the test masses has
certain complications.  This  is due to two reasons. The  first one is
that  magnetometers   usually  back-act   due  to   their  measurement
principles (i.e., they  also create their own  magnetic fields), while
the second is that the sensors selected for LISA Pathfinder have a
large  size,  which  conflicts  with space  resolution  and  with  the
possibility of having a sufficient number  of them to properly map the
magnetic field around the  test masses.  However, high-sensitivity and
small-size sensors that significantly  mitigate the two aforementioned
limitations exist, and have been  proposed to overcome these problems.
Thus,  these  sensors  will  be   likely  selected  for  the  magnetic
diagnostics  subsystem  of  eLISA.   Here we  perform  a  quantitative
analysis of the new magnetic  subsystem, as it is currently conceived,
and assess  the feasibility  of selecting these  sensors in  the final
configuration of the magnetic diagnostic subsystem.
\end{abstract}


\section{Introduction}
    
The eLISA  mission concept is  a proposed spaceborne gravitational wave
observatory  for the L3  theme ``The gravitational Universe''
(ESA)\,\cite{bib:L3}. The main purpose is the  study of the gravitational Universe in
the  frequency interval  between $0.1\,{\rm  mHz}$ and  $1\,{\rm Hz}$.
The  eLISA  concept   is  based  on  three   drag-free  spacecraft  in
one-million-kilometer  side equilateral  triangle.  Each  arm forms  a
laser  interferometer  between  free-falling  bodies  (46-mm-side
gold-platinum  cubes) to  measure  the weak  deformation of  spacetime
along   one    arm   of    the   interferometer   relative    to   the
other\,\cite{whitepaper}.   Due  to  the extremely  low  amplitude  of
gravitational waves\,\cite{bib:GWsources}, the test masses (TMs) are required to be shielded
from non-gravitational forces, which would disturb their pure geodesic
motion.  Consequently, environmental conditions around the TMs need to
be under stringent control, otherwise the different noise disturbances
would prevent the detection of gravitational waves.

The  eLISA noise  requirement  in terms  of free-fall accuracy is 
$3\,{\rm   fm\,s^{-2}  Hz^{-1/2}}$  per TM down  to
0.1\,{\rm mHz}\,\cite{whitepaper}. At frequencies below 1  mHz, the noise is dominated by
the residual acceleration noise caused by environmental effects, e.g.,
thermal,         magnetic          and         random         charging
fluctuations\,\cite{bib:LISAReq}.   Among   them,  one  of   the  main
contributors to the total acceleration noise budget is the surrounding
magnetic  field  in  the  spacecraft,   which  is  mostly  created  by
electronic units and  other components such as  the micro-thrusters of
the  satellite. The  magnetic field  and magnetic  field gradient  can
cause  a  non-gravitational  force  on  the TM  due  to  its  non-zero
magnetization ${\bf M}$ and susceptibility $\chi$. This spurious force
on the TM volume $V$ induced by a magnetic disturbance is given by:

\begin{equation}\label{equ:MagForce}
 {\bf F} = \left\langle\left[\left({\bf M} + 
           \frac{\chi}{\mu_0}\,{\bf B}\right)\Cdot\Nabla\right]{\bf B}
           \right\rangle V. 
\end{equation}

\begin{figure}[ht!]
\centering
\includegraphics[scale=0.4]{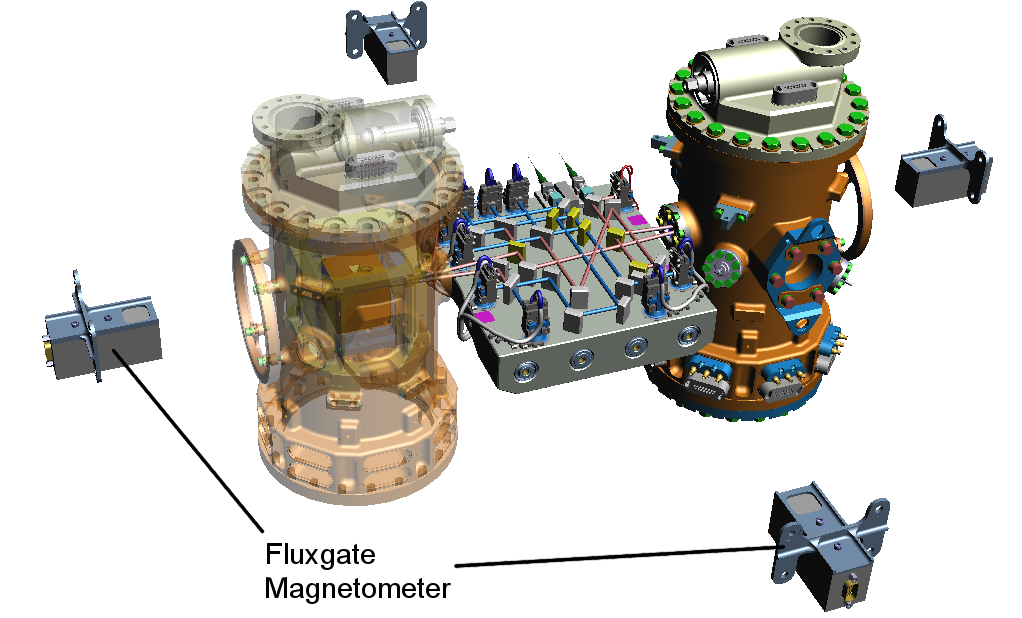}
\caption{The  payload  of LISA  Pathfinder,  with  the four  tri-axial
  fluxgate  magnetometers.   Each  of the  electrode  housings  (cubic
  structures) inside the vacuum enclosure (the two cylindrical towers)
  encloses one TM at its center (solid gold cube).}
\label{fig:LTP}
\end{figure}

While the magnetic  properties of the TMs (${\bf M}$  and $\chi$ ) are
known     owing     to     several     on-ground     and     in-flight
experiments\,\cite{bib:TMproperties, bib:InflightTM},  the  magnetic  field  environment
(${\bf  B}$ and  $\Nabla{\bf  B}$) at  the TM  locations  needs to  be
carefully evaluated during the mission.   To that end, eLISA will have
a set of magnetic sensors placed in key locations, with the purpose of
discerning  the   magnetic  noise   contributions  from   the  overall
acceleration  noise budget.   The ongoing research concerning the possible design of a magnetic  diagnostics subsystem  for
eLISA is based  on the experience with its precursor mission,  LISA Pathfinder, in
which high-performance  fluxgate magnetometers were chosen  because of
their sensitivity  and availability  for space  applications\,\cite{bib:LPFDiagnostics, LPFMag}. However,
these  sensors  are  bulky  ($94\,{\rm  cm^3}$)  and  have  a  large
ferromagnetic sensor head ($\sim2\,{\rm  cm}$ long). These reasons led
to placing only four tri-axial  sensors at somewhat large distances from
the    TMs    ($\geq18.85\,{\rm    cm}$)    to    avoid    back-action
disturbances. Besides, the size of the sensor head also conflicts with
the space  resolution, which might be  another source of error  in the
determination  of the  magnetic  field.  A  view  of the  magnetometer
location   in    the   LISA    Pathfinder   payload   is    shown   in
figure\,\ref{fig:LTP}.

We stress that unlike critical drag-free technology that needs from the in-flight experiments to be fully proved, the feasibility of the magnetic measurement subsystem can be verified in depth from the analysis of the ground test campaigns. On the basis of the previous analysis for LISA  Pathfinder, the selected  arrangement  of magnetic  sensors
resulted in an unsatisfactory estimation  of the magnetic field in the
TM  region   using  classical  interpolation   methods.   Accordingly,
alternative  approaches  needed  to  be adopted.   In  particular,  an
interpolation scheme based  on Neural Networks needed  to be developed
\cite{bib:Neural}.  For  the case  of eLISA, a  more robust  method to
reconstruct the magnetic field at the position of the TMs is foreseen.
This  requires a  sufficient  number of  smaller magnetometers,  which
additionally must be placed closer to  the TMs. Besides, it is required
that back-action effects should be negligible.  All this motivated the
study  of  alternatives   to  fluxgate  magnetometers.   Specifically,
magnetoresistances\,\cite{bib:AMRTemp} or chip-scale atomic vapor cell
devices\,\cite{BudkerBook} have been proposed.  These high-sensitivity
and  small   sensors  will  significantly  mitigate   the  limitations
mentioned above. Thus, they will be  likely chosen to be integrated in
the magnetic diagnostics subsystem in  eLISA, improving the quality of
magnetic field interpolation.   

All in all, the LISA Pathfinder magnetic diagnostics is fully integrated in the spacecraft due to launch in 2015, and the mission operations together with the data analysis are expected to be completed by 2016. Regarding the magnetic interpolation process to be used in LISA Pathfinder, the aforementioned Neural Networks algorithms is at the present the most promising one, although it is still an ongoing activity. On the other hand, eLISA is currently under the mission concept study and the critical technologies need to be be available for the mission concept selection in 2020. The reader will find details in\,\cite{bib:LPFstatus} about the general status of eLISA and its precursor LISA Pathfinder.

In this paper we study assess the feasibility of using Anisotropic  Magnetoresistance  sensors   (AMRs)  for  estimating  the
magnetic field and its gradient at the location of the TMs. The paper
is  organized   as  follows.   In   section\,\ref{sec:interpolation} 
the
theoretical  methods   for  the   magnetic  field   interpolation  are
explained, while  in section\,\ref{sec:layout}  the sensor array  and
the
distribution of  the magnetic sources  are addressed.  The  results of
our analysis  are presented  in section\,\ref{sec:results}.   Finally,
we
draw our conclusions in section\,\ref{sec:conclusions}.


\section{Interpolation methods}
\label{sec:interpolation}

The magnetic  field at the TM  location must be inferred  according to
the information given by the  magnetometer readings. We are interested
in  a robust  method  that  works without  previous  knowledge of  the
spacecraft magnetic  field environment.  The reasons  for   this  choice   are  that  the
expected local spacecraft field might be affected by  possible changes of the
magnetic characteristics of the spacecraft during launch or during
the  lifetime  of  the  mission,  by  deviations  from  the  on-ground
performance, and by varying operational modes in the spacecraft.  Hence,
methods making use of {\em a priori} knowledge, such  as neural networks 
or Bayesian frameworks that  yield remarkable  results in similar 
estimation problems\,\cite{bib:Neural,bib:MCMC,bib:Bayesian} will  not be considered
here. Instead, in this  work we adopt as our interpolation
tool  the multipole  expansion technique based only on the magnetometer readouts.  The results  obtained using
this method are then compared  with other theoretical approaches, such
as the Taylor series and the distance weighting interpolating methods. 
In  the  following  sections  we briefly  describe  the  interpolation
methods employed for this study.

\subsection{Multipole expansion}

Since the magnetic sources in the  spacecraft are located far from the
origin of the  coordinate system (chosen at the centre  of the TM) and
assuming  the  material inside  the  vacuum  enclosure is  basically
non-magnetic, the magnetic  field in this region can  be considered to
be essentially a  {\em vacuum} field ($\Nabla \times {\bf  B} = \Nabla
\cdot  {\bf B}  = 0  $).  Hence,  the estimated  magnetic field  ${\bf
B}_{\rm e}$ obtained employing an array  of $N$ sensors can be written
as  the general  solution to  Laplace's equation  centered at  the TM,
which  can  be  expressed  in  terms  of  an  expansion  in  spherical
harmonics:

\begin{equation}
 {\bf B}_{\rm e}({\bf x}) = \Nabla\Psi({\bf x}) = \sum_{l=1}^L\sum_{m=-l}^l\,
 M_{lm}(t)\,\Nabla[r^l\,Y_{lm}({\bf n})],
 \label{eq:multipole}
\end{equation}

\noindent where $r \equiv |{\bf x}|$ and ${\bf n \equiv x} /r$ are the
spherical coordinates of the field at ${\bf x}$. $M_{lm}$ and $Y_{lm}$
are the multipole coefficients and the standard spherical harmonics of
degree $l$ and order $m$, respectively\,\cite{bib:jackson}.

The accuracy of  the estimation of the magnetic field  is given by the
order  of the  expansion, which  depends  on the  number of  multipole
coefficients that can  be computed. Specifically, the  accuracy of the
interpolation  is  given  by  the   number  of  known  magnetic  field
measurements at the  boundary of the volume where  the field equations
are considered.  In our  case these measurements  are provided  by the
number     of    magnetometers     placed    in     the    spacecraft.
Table\,\ref{tab:multipole} shows  the minimum number  of magnetometers
required to model  the magnetic field with a second,  third and fourth
order multipole expansion.

\begin{table}[t!]
\caption{Order  of  the  multipole   expansion,  number  of  multipole
  coefficients and number of needed magnetometers. The number of triaxial magnetometers (last column) necessary to achieve the desired order satisfies the condition $3 \cdot N \geq  L (L + 2)$.
\label{tab:multipole}}
\begin{indented}
\lineup
\item[]\begin{tabular}{@{}clcc}
\br
      \textbf{Expansion}    & \textbf{Equivalent}         &  \textbf{\# of $M_{lm}$}    &\textbf{\# of triaxial}  \\ 
      \textbf{order}        & \textbf{multipole}          &  \textbf{ coefficients}     &\textbf{magnetometers}   \\ 
      \textbf{L}            &                             &  \textbf{[L (L + 2)]}       &\textbf{[N]}             \\ 
\mr
      2                     &     Quadrupole              &   \08                       &  3                      \\
      3                     &     Octupole                &    15                       &  5                      \\
      4                     &     Hexadecapole            &    24                       &  8                      \\
\br
\end{tabular}
\end{indented}
\end{table}
  
The  coefficients  $M_{lm}$  are  found  by  minimizing  the  equation
$\partial\varepsilon^2/\partial M_{lm} = 0$, where the square error is
defined as
\begin{equation}
 \varepsilon^2(M_{lm}) = \sum_{s=1}^{N}\,\left|
 {\bf B}_{\rm m}({\bf x}_s) -
 {\bf B}_{\rm e}({\bf x}_s)\right|^2,
 \label{eq.12}
\end{equation}
${\bf B}_{\rm m}$ is the readout of the triaxial magnetometer, and $N$
is  the  total number  of  magnetometers.  This  is done  employing  a
least-squares  method. Once  the system  of equations  is solved,  the
computed    coefficients    $M_{lm}$     can    be    inserted    into
equation~(\ref{eq:multipole}), replacing the magnetometer's position,
${\bf
x}_s$, by the  TM position, ${\bf x}_{\rm TM}$, to  finally obtain the
value of the interpolated field at the TM location.

\subsection{Taylor series}

The magnetic  field at the TM  position inferred from the  readings of
the magnetometers can  also be approximated by a  Taylor expansion. As
in the case in which the multipole expansion is employed, the order of
the Taylor  series is  determined by the  number of  magnetometer data
channels. In this  case the magnetic field at the  position of the TMs
can be approximated by the following expression:

\begin{equation}
 {\bf B_{\rm m}}({\bf x}_s) = {\bf B_{\rm e}}({\bf x_{\rm TM}}) +
 \sum_{n=1}^L \sum_{i=1}^3\frac{\partial^{n}{\bf B_{\rm e}}({\bf x}_{\rm TM})}{\partial x_i}\frac{(x_{s,i}-x_{{\rm TM},i})^{n}}{n!},
 \label{eq:taylor}
\end{equation}

\noindent where the origin of coordinates  is defined at the centre of
the  respective TM  (${\bf x}_{\rm  TM}$), and  ${\bf x}_{s}$  are the
magnetometer  locations.  ${\bf  B_{\rm  e}}({\bf  x_{\rm  TM}})$  and
$\partial^{n}{\bf  B_{\rm e}}\,({\bf  x}_{\rm  TM})/\partial x_i$  are
calculated considering that the magnetic  field around the TM has both
zero  divergence and  curl, i.e.  the magnetic  field gradient  tensor
$\nabla^n{\bf B}$  is a symmetric  and traceless matrix. Thus,  only a
total of 5 independent components need to be computed.

\subsection{Distance weighting}

This method consists  in computing the field as a  weighted sum of the
different  magnetometer  readings.  The calculation  is  performed  as
follows:

\begin{equation}
{\bf B}_{\rm e} = \sum_{s=1}^N a_s {\bf B_{\rm m}}({\bf x}_s),
\end{equation}

\noindent where ${\bf  B_{\rm m}}({\bf x}_s)$ are the  readouts of the
magnetometers. The weighting factors $a_s$ are given by:

\begin{equation}
a_s = \frac{1/r_s^n}{\sum_{i=1}^{N} 1/r_i^n},
\label{eq:as}
\end{equation}

\noindent where $n$ specifies the order of the interpolation and $r_i$
are  the distances  between  the  point at  which  the  field must  be
estimated and the specified magnetometer.

      
\section{Magnetic sources and sensor layout}
\label{sec:layout}

We  first note  that the  interplanetary DC  field is  expected to  be
more than one order of magnitude weaker than  the sources of magnetic field present
inside  the  spacecraft\,\cite{bib:Bip1}. By  design,  there  are  not any  sources  of
magnetic  field  inside  the  vacuum enclosure  cylinder.   Since  the
distribution of the different subsystems in eLISA is not fully defined
yet, the distribution of the magnetic sources in the spacecraft is not
known. However, in order to provide a realistic scenario to assess the
performance  of  our  proposed  interpolation  methods,  we  make  the
following assumptions. We first assume that the magnitude and location
of the  magnetic sources  are the ones  measured for  LISA Pathfinder.
Moreover, we  also assume that  the sources  of magnetic field  can be
modeled as point  magnetic dipoles. With these assumptions  a batch of
$10^3$ different  magnetic realizations  is generated using  the fixed
locations and  magnitudes of  the magnetic field  of the  sources, but
with orientations randomly drawn according to normal distributions for
each of the components.
    
\begin{figure}[t!]
\centering
\includegraphics[width=0.5\columnwidth]{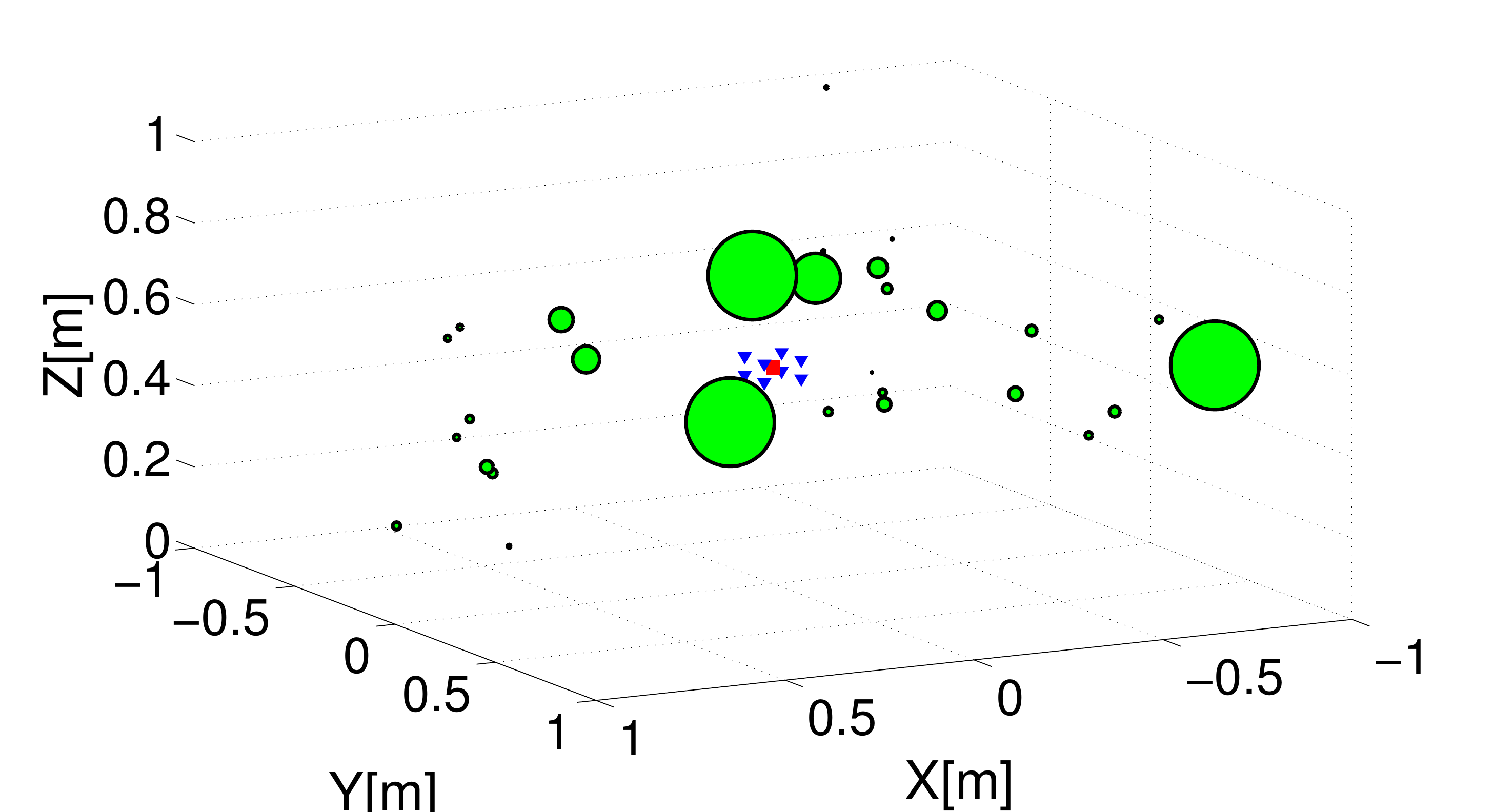}
\includegraphics[width=0.4\columnwidth]{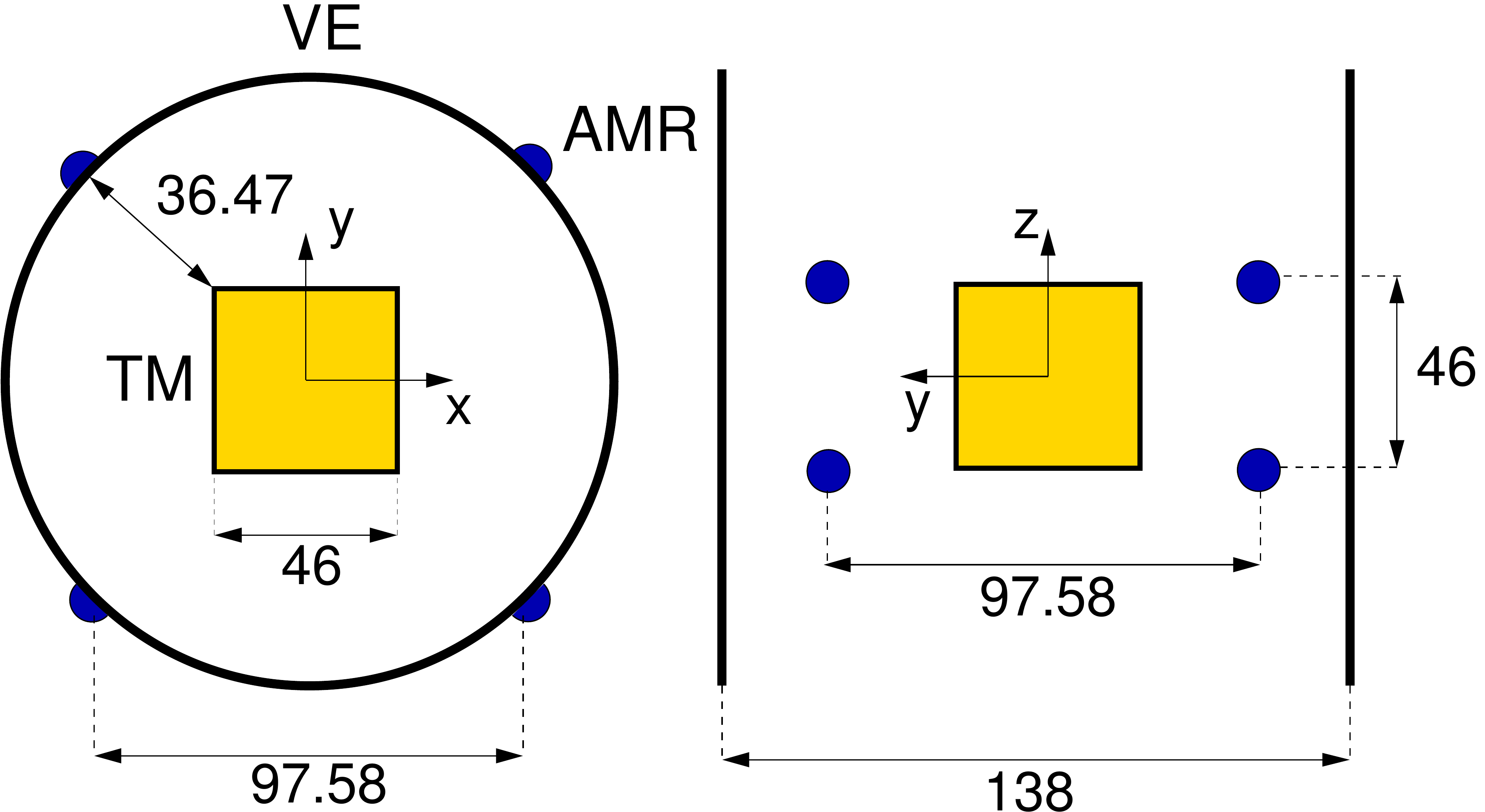}
\caption{Left:  A view  of  the 29  measured  dipole magnetic  sources
  (green dots: the size is proportional to their magnetic moment), the
  test  mass   (red  square)  and   the  8  AMR   magnetometers  (blue
  triangles).  Right:   Sensor  array  configuration  on   the  vacuum
  enclosure (Units in mm).}
\label{fig:conf}
\end{figure}

The  adequate  location  and  number of  magnetometers  stem  from  a
trade-off between the  accuracy of the reconstruction  of the magnetic
field map and the magnetic  disturbances generated by the magnetometer
itself  on the  TM region.  In  order to  quantify the  effect of  the
sensors,  the magnetic  moment  of an  AMR has  been  measured with  a
Superconducting  Quantum  Interference  Device (SQUID)  for  different
configurations. Our analysis based on the SQUID measurements shows that symmetrical placements with
four and eight sensors are the  preferred options in order to minimize
the magnetic  back-action effects. Moreover, when  eight   sensors 
are allocated in  a symmetrical configuration  on the walls of  the
vacuum enclosure  their contribution  to the  magnetic  budget
is negligible~\cite{bib:AMRBackAction, AMR_LISA}.  Figure\,\ref{fig:conf}
displays the distribution  of the
sources of  the magnetic field  in the LISA Pathfinder  spacecraft and
the 8-sensor  layout that is being considered in the current analysis for eLISA.
Additionally, we  carried out  noise measurements of  the magnetometer
with the  sensor allocated  inside a magnetic  shield, and  obtained a
noise floor of  $\sim 150\,{\rm pT Hz^{-1/2}}$\,\cite{bib:AMRTemp}. 
Accordingly, to mimic the electronic noise of  the system, this noise
is added  to the simulated readouts of the magnetometers.
   
Finally,  in  order   to  assess  the  performance  of   each  of  the
interpolating  methods, the  interpolated magnetic  field is  compared
with  the exact  one, assuming  that  the different  magnetic sources behave  as
point dipoles.  Then,  the  total   magnetic  field
generated by the sources can be calculated as: 
  
\begin{equation}
 {\bf B}({\bf x})= \frac{\mu_{0}}{4\pi}\sum^{n}_{a=1}\frac{3({\bf
m}_{a}\Cdot {\bf n}_{a}){\bf n}_{a}-{\bf m}_{a}}{|{\bf x}-{\bf x}_{a}|^{3}},
\label{eq:fieldB}
\end{equation} 

\noindent  where  ${\bf  m}_{a}$  are  the  magnetic  dipolar  moments
measured for  the different subsystems,  ${\bf x}_{a}$ are  the source
positions  and  $n$  is  the   number  of  sources.  The  corresponding
expression for the magnetic field gradient is:

\begin{eqnarray}
\fl \frac{\partial B_i}{\partial x_j} =\frac{\mu_{0}}{4\pi}\sum^{8}_{a=1}\frac{3}
 {|{\bf x}-{\bf x}_{a}|^{4}}[(m_{a,i}n_{a,j}+m_{a,j}n_{a,i}) + ({\bf m}_{a}{\bf \Cdot}{\bf n}_{a})(\delta_{ij}-5n_{a,i}n_{a,j})],
 \label{gradB}
\end{eqnarray}

\noindent where $\delta_{ij}$ is Kronecker's delta.


\section{Results}
\label{sec:results}

\subsection{Magnetic field reconstruction}   

\begin{figure}[t!]
\centering
\includegraphics[width=0.48\columnwidth]{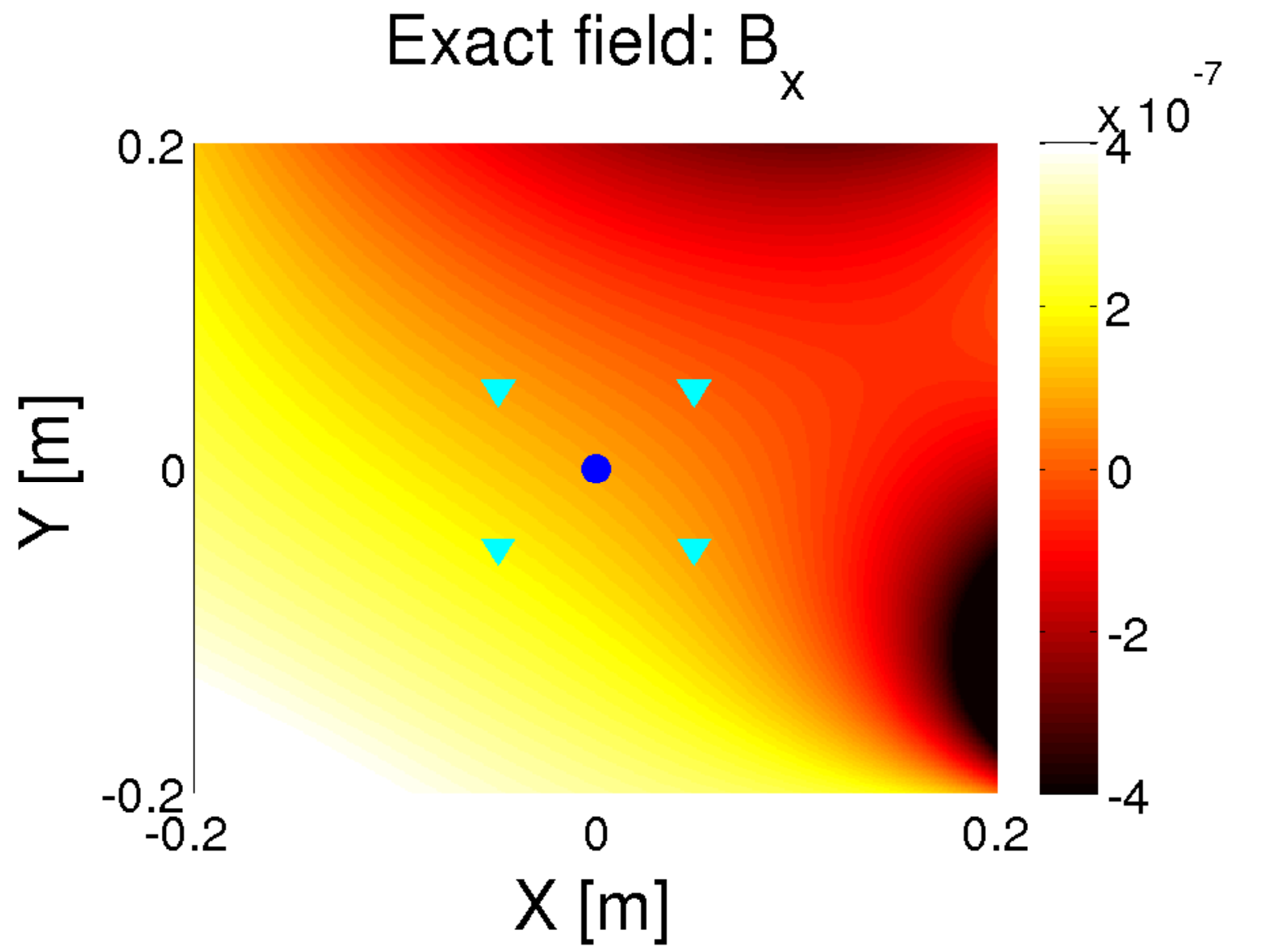}
\includegraphics[width=0.48\columnwidth]{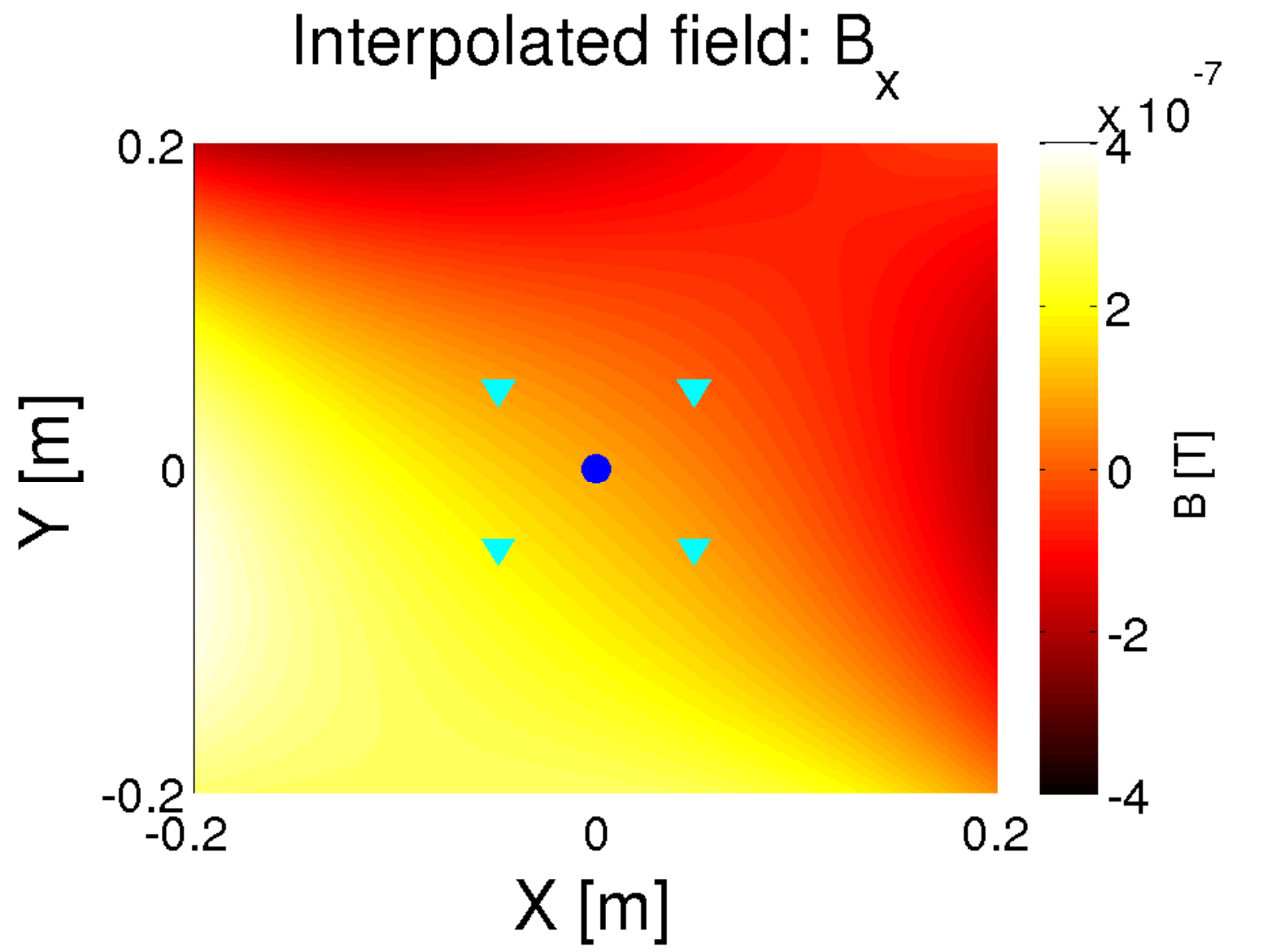}
\caption{Contour plot  of the  exact (left) and  reconstructed (right)
  magnetic field $B_x$  for a given source  dipole configuration using
  multipole expansion  with 8  magnetometers. The  positions of  the 8
  magnetometers (cyan  triangles) and of  the test mass  (blue circle)
  are also represented.}
\label{fig:ylm8}
\end{figure}

As  previously   explained,  to   validate  the  performance   of  the
reconstruction algorithm,  a batch of dipoles  with randomly generated
orientations were simulated and the  exact magnetic field for each one
of these realizations was compared with the interpolated results. The
left
panel  of  figure\,\ref{fig:ylm8}  shows   the  $x$-component  of  the
magnetic field map produced by one of these random configurations. The
results are then compared in the right panel with those obtained using
one  of our  interpolating methods,  in this  case the  magnetic field
reconstructed    using    multipole    expansion.     As    seen    in
section\,\ref{sec:interpolation},  a  multipole expansion based
only on eight triaxial magnetometers readings is able to resolve the
magnetic field up to the
hexadecapole     structure,    by     computing     24    terms     in
equation~(\ref{eq:multipole}).  Overall,  the field  qualitatively
resembles
the exact  one, although there  are apparent differences far  from the
positions  of  the  TMs.  However,   note  that  the  success  of  the
reconstruction method  is determined by  the accuracy achieved  at the
region  of interest,  i.e.  at the  TM locations.  We  perform a  more
quantitative analysis for the three components of the field below.

\begin{figure}[t!]
\centering
\includegraphics[width=0.48\columnwidth]{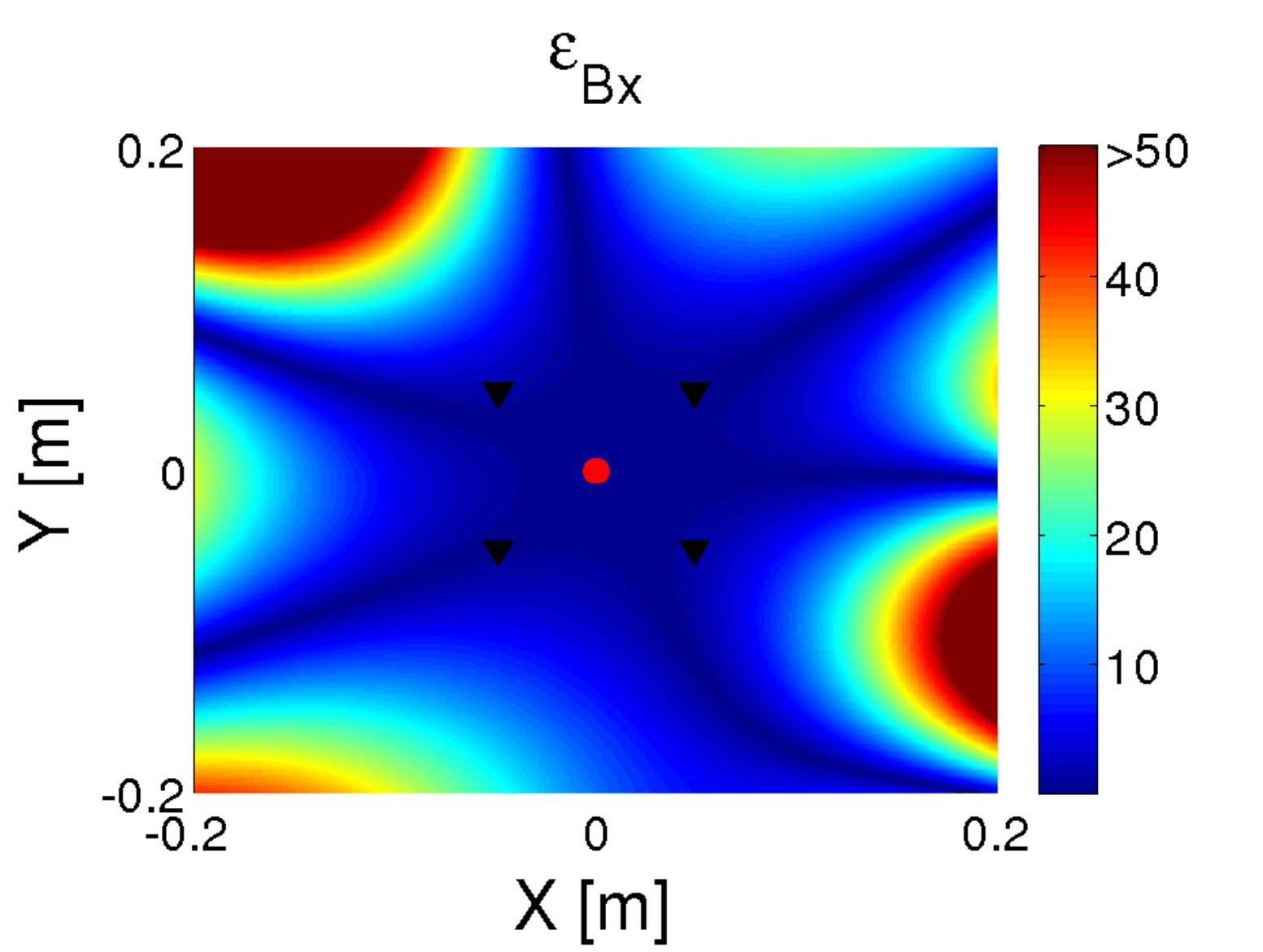}
\includegraphics[width=0.48\columnwidth]{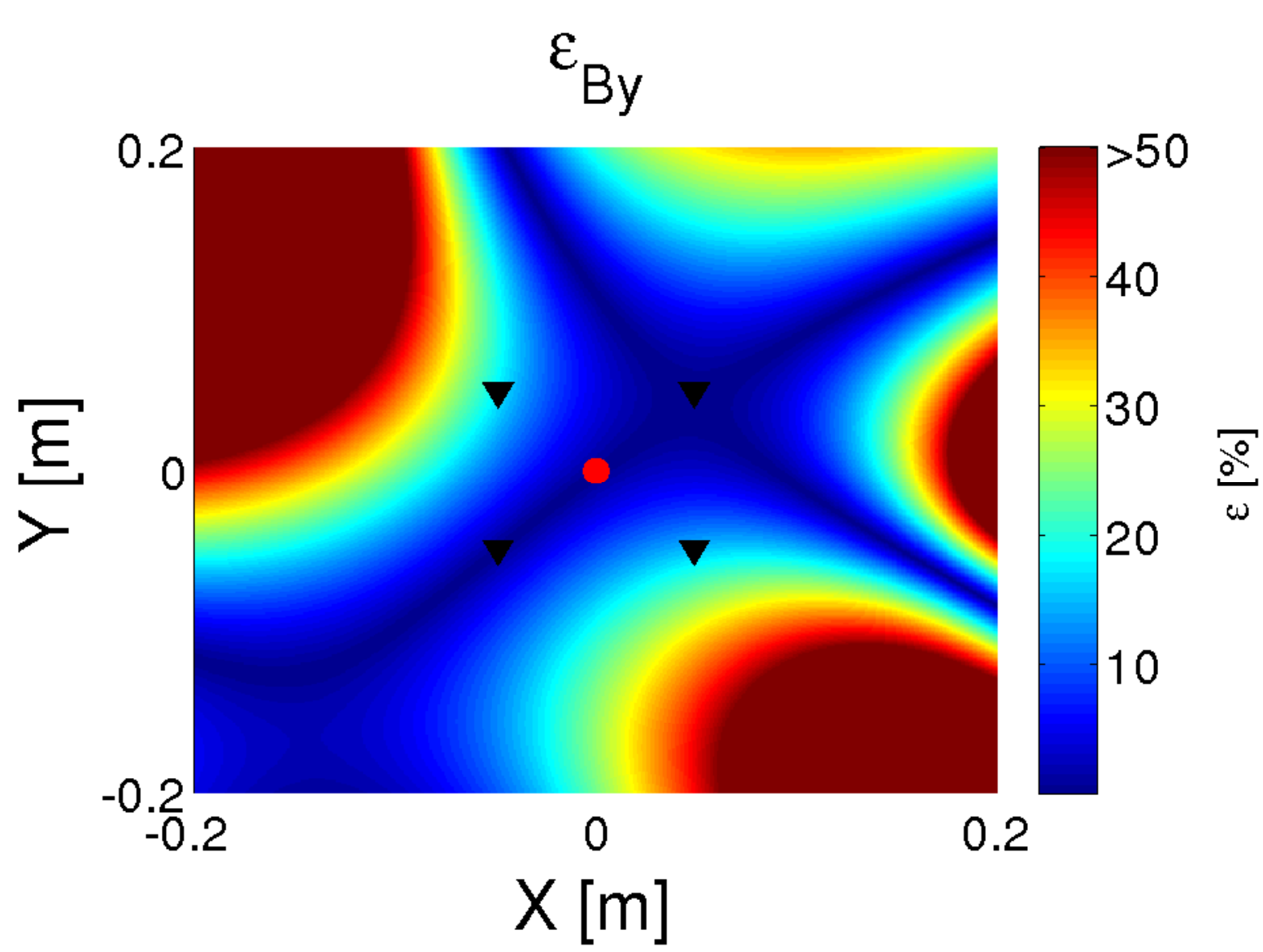}\\
\includegraphics[width=0.48\columnwidth]{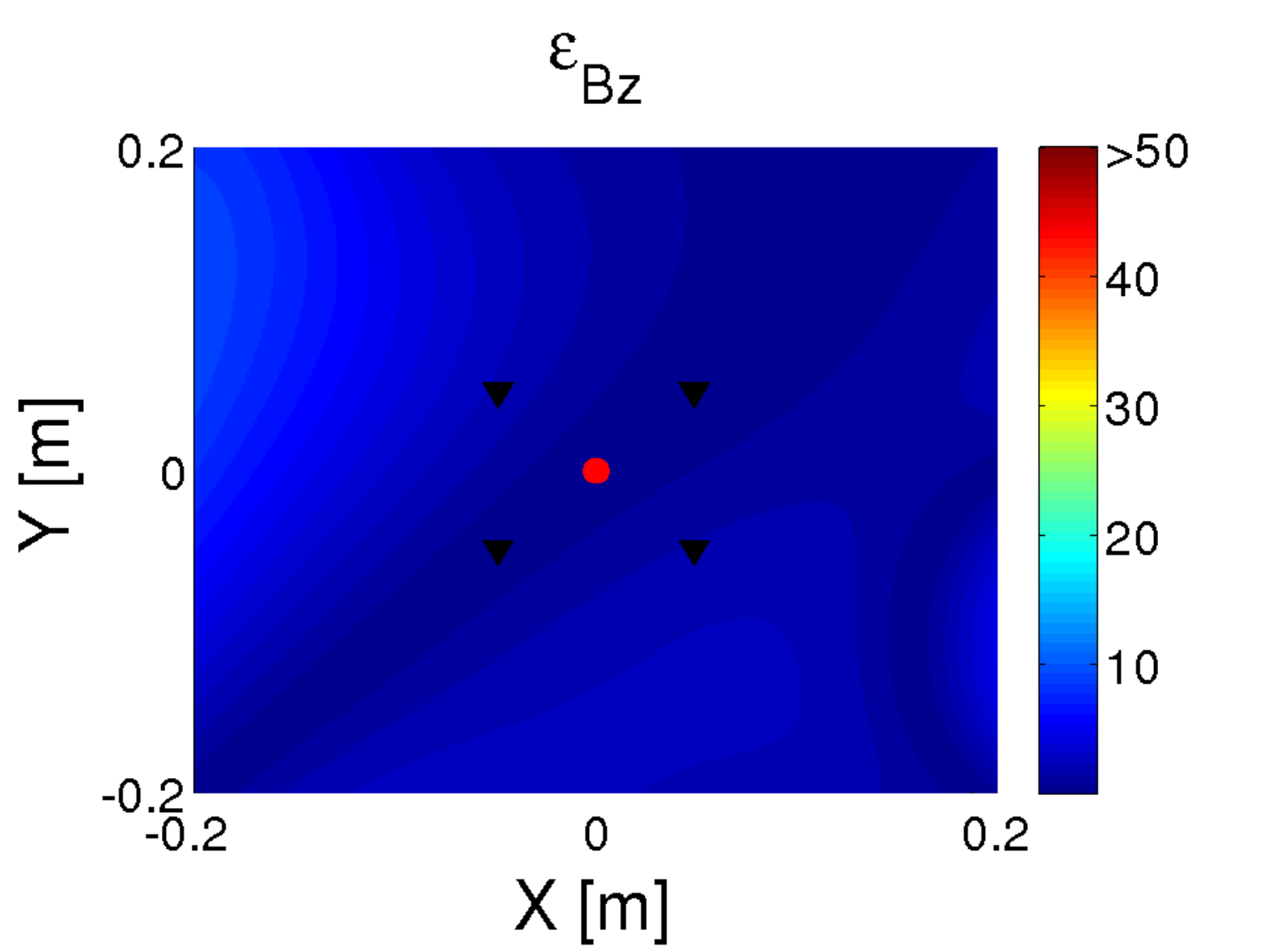}
\includegraphics[width=0.48\columnwidth]{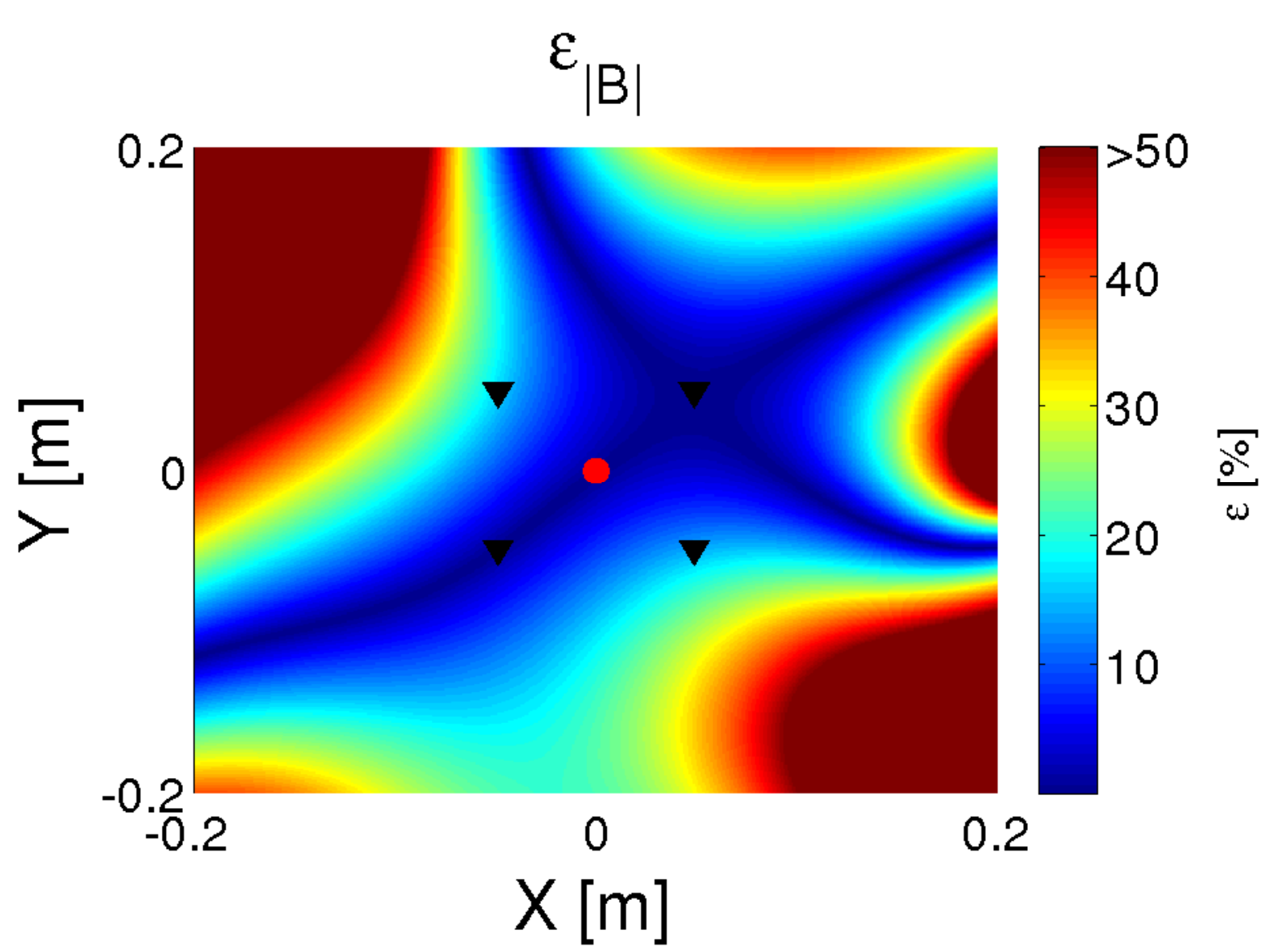}
\caption{Relative  errors  in the  estimation  of  the magnetic  field
  components and  the modulus.   To calculate  the relative  error for
  each field component,  the absolute error is divided  by the modulus
  of the  exact value  in order  to avoid infinities  when one  of the
  vector components is close to zero $\varepsilon_{B{\rm x}} = (B_{\rm
  r,x} - B_{\rm e,x})/|{\bf B}_{\rm r}|$.}
\label{fig:errorylm8}
\end{figure}

The differences (in percentage) between the interpolated field and the
source dipole  model field  are shown  in figure\,\ref{fig:errorylm8}.
Contour  plots for  the  three  components and  the  modulus show  the
accuracy achieved by the multipole algorithm.   As can be seen in this
figure, the smallest  differences occur in the region  enclosed by the
magnetometers.  Moreover, the accuracy of the interpolating algorithm is
good in  the central area  of the electrode  housing, where the  TM is
located.

\begin{figure}[t!]
\centering
\includegraphics[width=1\textwidth]{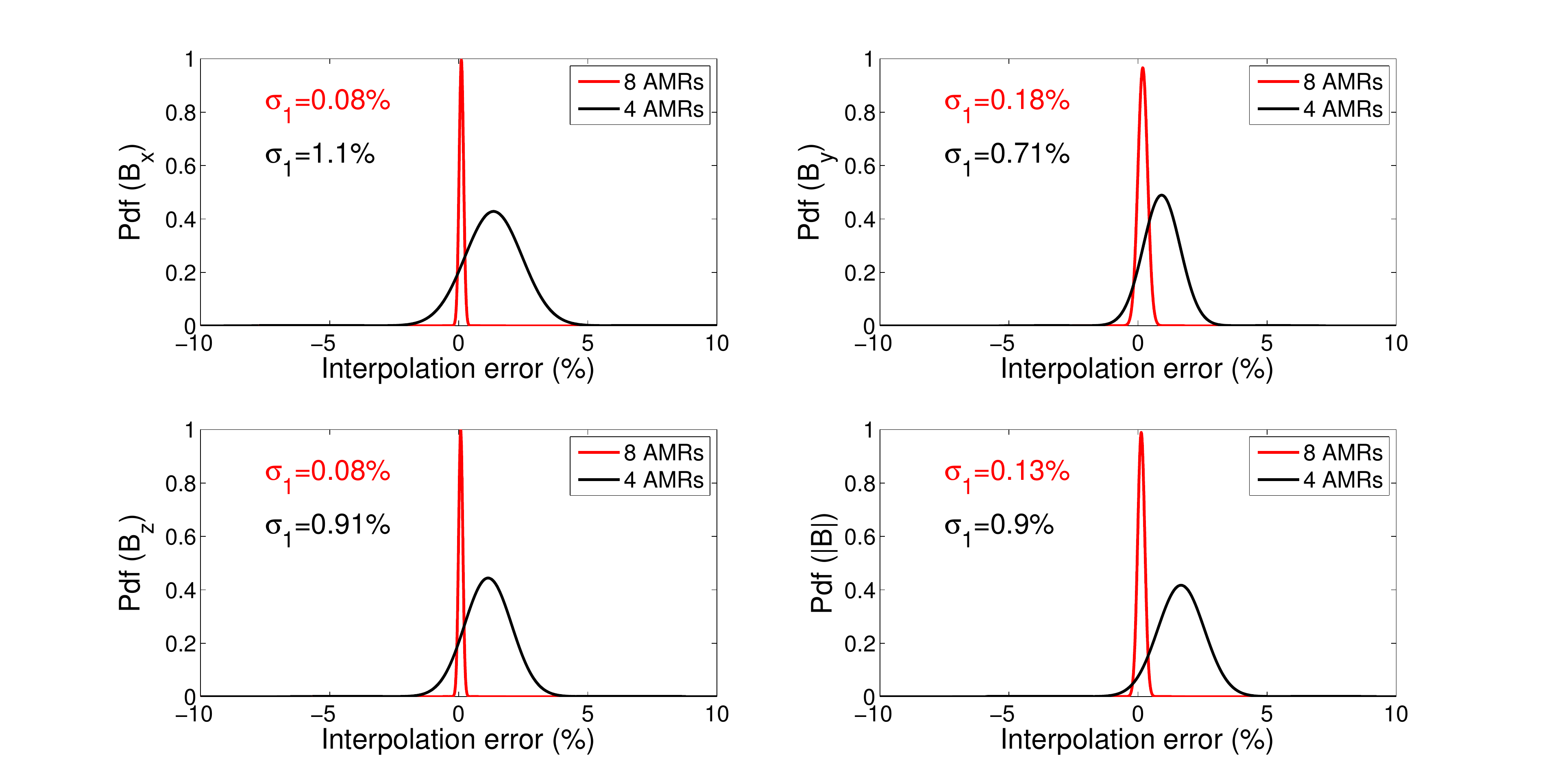}
\caption{Distributions of the  relative errors at the  TM position for
  $N  = 10^3$  random  cases  for four  (black)  and  eight (red)  AMR
  sensors.}
\label{fig:histogramB}
\end{figure}

To  further confirm  the  validity and  general  applicability of  the
multipole   expansion  we   compared  the   differences  between   the
interpolated and  exact magnetic field at  the position of the  TM for
three  different sensor  layouts. Specifically,  we first  adopted the
LISA Pathfinder configuration.  In  this layout fluxgate magnetometers
are used, as  depicted in figure\,\ref{fig:LTP}.  In a  second step we
did the same adopting four AMRs  placed around the vacuum enclosure at
the height of  the electrode housing center.  Finally,  we carried out
the same  calculation this  time adopting  eight AMRs,  as graphically
displayed in figure\,\ref{fig:conf}. Average  and maximum field errors
relative  to  the   modulus  ($\overline{\varepsilon}_{\bf  |B|}$  and
${\varepsilon}_{{\bf  |B|},{\rm max}}$)  and to  the field  components
($\overline{\varepsilon}_{B_i}$) over the $10^3$ random runs are shown
in table\,\ref{tab:errors}.  In the  LISA Pathfinder configuration the
accuracy of  the reconstructed field  at the  TM is poor  and presents
large variations when the multipole  expansion is used. In particular,
the estimation errors can be as high as $737 \%$.  This is the natural
consequence of  having placed the sensors  too far from the  center of
the TM.  Instead,  when AMRs are used, the sensors  can be placed much
closer to the center of the TM,  due to its smaller size and intrinsic
magnetic moment.  The results when the same number of magnetometers is
employed  show significant  improvements,  with maximum  errors up  to
$15\%$.  Finally,  the estimation  errors are reduced  by a  factor of
$\sim 6$  (${\varepsilon}_{{\bf |B|},{\rm  max}} = 2.4\%$)  when eight
sensors  are used.  In this  case  the hexadecapole  expansion can  be
employed, and this obviously results  in an improved performance of the
interpolating algorithm.  Last, in figure\,\ref{fig:histogramB} the
distribution of the estimation errors for the randomly simulated cases
is shown.  This figure clearly  shows that the standard deviations are
$\leq  1.1\%$ and  $  \leq 0.18\%$  for the  4-AMR  and 8-AMR  layouts,
respectively.   This  proves  that   the  averaged  estimation  errors
($\overline{\varepsilon}_{\bf |B|}  \leq 0.2\%$)  are robust  and that
the performance  of   the  multipole  interpolating  algorithm   is  good,
providing  reliable estimated  values  of the  magnetic  field at  the
location of the TM.

\begin{table}[t!]
 \caption{Relative  errors of  the  magnetic field  estimation at  the
   positions  of   the  TM.  $\overline{\varepsilon}_{\bf   |B|}$  and
   $\overline{\varepsilon}_{B_i}$ are  the mean  error for a  batch of
   $10^3$ randomly orientated magnetic sources relative to the modulus
   ${\bf |B|}$ and  to the field component  ${B_i}$, respectively. The
   denominator  in $\overline{\varepsilon}_{B_i}$  is  closer to  zero
   than that  of the modulus $\overline{\varepsilon}_{\bf  |B|}$, this
   translates  in larger  errors for  the $x$-component  than for  the
   modulus. \label{tab:errors}} 
\begin{indented}
\lineup
\item[]
\begin{tabular}{@{}lllllllllllll}
       \br
       {\bf Error}  &  \multicolumn{4}{c}{{\bf LPF} (4 Fluxgates)}  &  \multicolumn{4}{c}{{\bf eLISA} (4 AMRs)}   &         \multicolumn{4}{c}{{\bf eLISA} (8 AMRs)}  \\
         ($\%$)      & $B_x$ & $B_y$ & $B_z$ & $\left|{\bf B}\right|$ & $B_x$ & $B_y$ & $B_z$ & $\left|{\bf B}\right|$ & $B_x$ & $B_y$ & $B_z$ & $\left|{\bf B}\right|$\\
       \mr
       $\overline{\varepsilon}_{\bf |B|}$       & \038.2 &\028.1 & \020.9 & \032.5 &  \01.4 &  1   & \01.1 &  \01.8 & 0.1 &  0.2 &  0.1 &  0.1\\
       ${\varepsilon}_{{\bf |B|},{\rm max}}$    &737.7   & 340.3 &  327.6 &  803.2 &  15.0  &  7.7 &  14.0 &   13.3 & 0.9 &  2.4 &  1.4 &  2.0\\
       $\overline{\varepsilon}_{B_i}$           & 697.9  & 202.1 &  184.5 & \032.5 &  13.7  &  3.8 &  \07.8 & \01.8 & 0.6 &  0.8 &  5.3 &  0.1\\       
       \br
    \end{tabular}
\end{indented}
\end{table}

The  results  described  so  far were  obtained  by using  the  multipole
expansion algorithm. However,  other  interpolation schemes were
detailed  in section\,\ref{sec:interpolation}, and their
performance were compared with that of the multipole expansion in
Table~\ref{tab:errorsAlgorithm}. The order of the interpolation in the
distance weighting  method is set to  $n =1$. Nevertheless, this
choice is not relevant  due to  the physical symmetry  of the  sensor
 placement, i.e.,  the distances  $r_s$,  and consequently  the weighting 
factors $a_s$, are  equivalent for  the eight  magnetometers.  For  the 
Taylor expansion, the second and the terms involving higher-order derivatives 
are  negligible  due  to the  symmetry  of  the  magnetic distribution. Thus,
the Taylor  approach mainly estimates the magnetic
field  as a  linear approximation.   For  this reason,  we expect  the
results of the interpolation to  be almost identical to those obtained
using the distance weighting method.  Table\,\ref{tab:errorsAlgorithm}
shows the accuracies of the estimation of the magnetic field at the
position of the TM for the three methods employed in this work. As can
be seen,  the multipole expansion outperforms  by far the rest  of the
methods described previously.

\begin{table}[t!]
 \caption{Maximum  errors  of  the  estimated magnetic  field  at  the
   position of the TM  using different interpolation methods, see text
   for details.}
\label{tab:errorsAlgorithm}
\begin{indented}
\lineup
\item[]
\begin{tabular}{@{}lllll}
       \br
       {\bf Error}  &  \multicolumn{4}{c}{${\varepsilon}_{{\bf |B|},{\rm max}}\, [\%]$ } \\
                    & $B_x$ & $B_y$ & $B_z$ & $\left|{\bf B}\right|$ \\
       \mr
       Distance weighting   & 8.0 & 4.0 & 7.7 & 7.9 \\
       Taylor expansion     & 8.0 & 4.0 & 7.7 & 7.9 \\
       Multipole expansion  & 0.9 & 2.4 & 1.4 & 2.0 \\
       \br
    \end{tabular}
\end{indented}
\end{table} 

\subsection{Reconstruction of the magnetic field gradient}

\begin{figure}[t!]
\centering
\includegraphics[width=0.48\columnwidth]{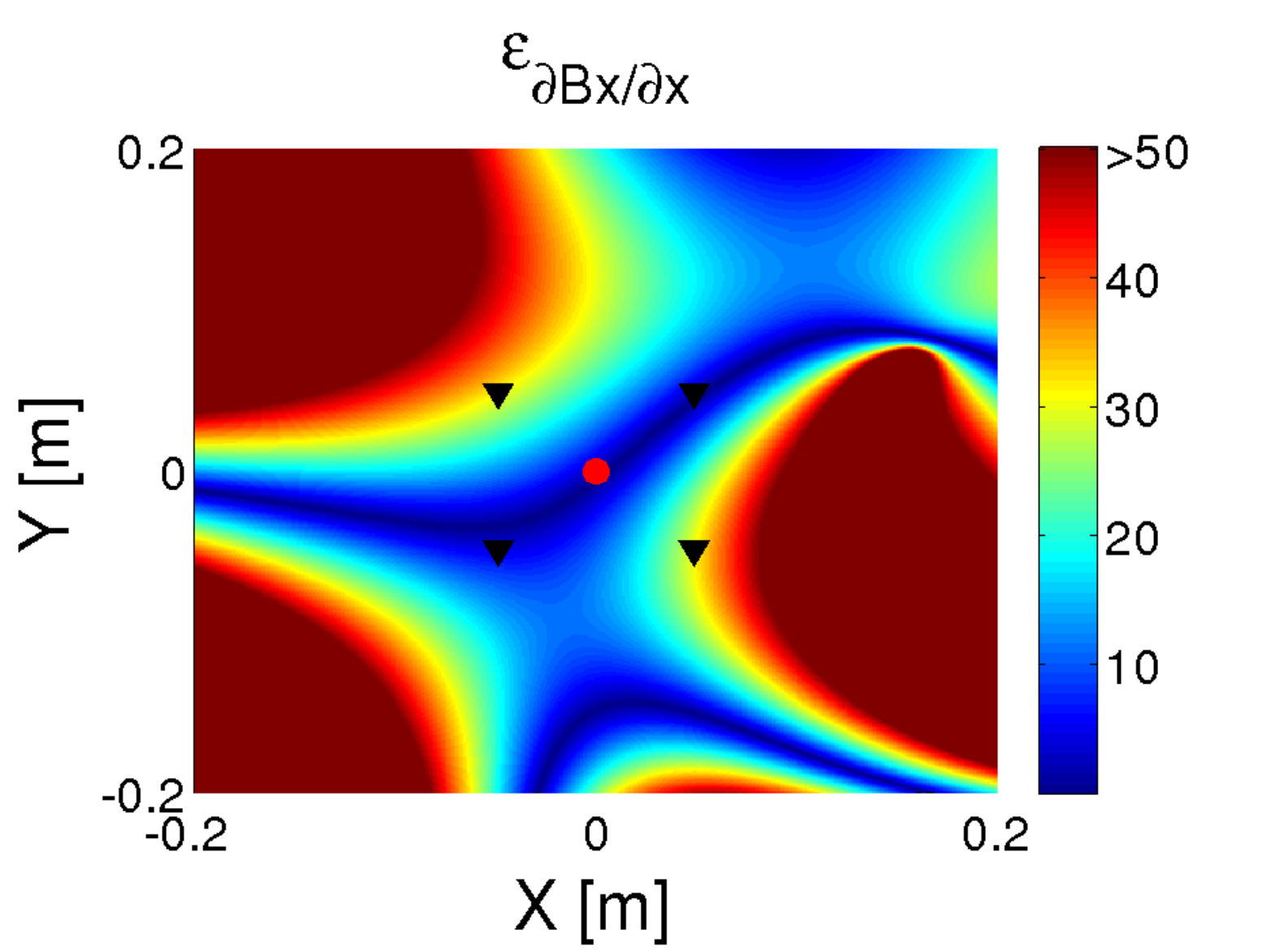}
\includegraphics[width=0.48\columnwidth]{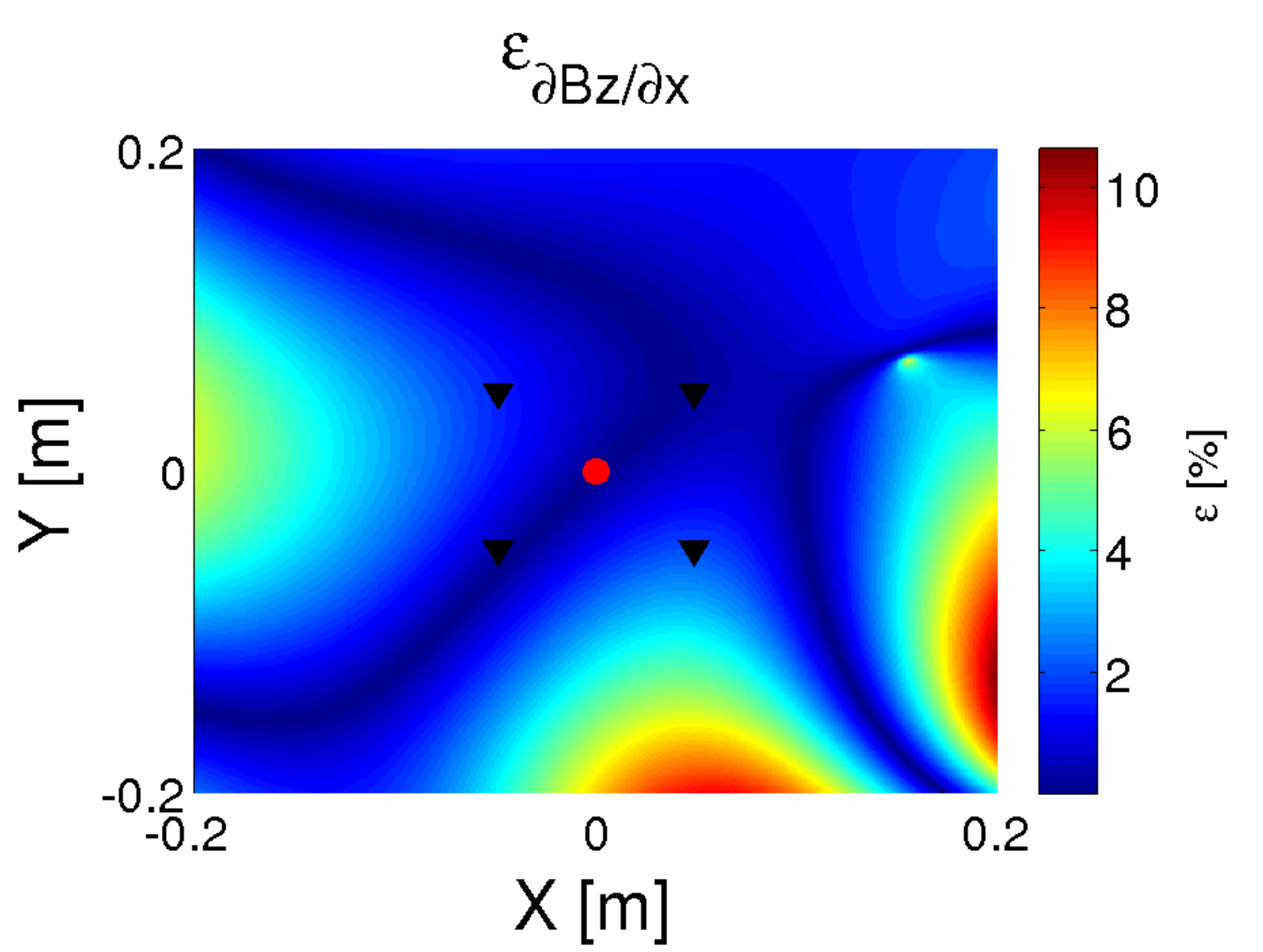}
\caption{Relative  errors  in the  estimation  of  the magnetic  field
  gradient.  Here,  for  the  sake   of  clarity,  we  only  show  two
  components, $\partial B_x/\partial x$ and $\partial B_y/\partial y$.
  The   relative   error   is   computed   as   $\varepsilon_{\partial
  B_{i}/\partial j}  = (\partial  B_{{\rm r},i}/\partial_j  - \partial
  B_{{\rm  e},i}/\partial_j)/|\partial {\bf B}_{{\rm  r}}/\partial_j|$. Note
  the different scale for the error bars.}
\label{fig:errorylm8Grad}
\end{figure}

\begin{figure}[ht!]
\centering
\includegraphics[scale=0.55]{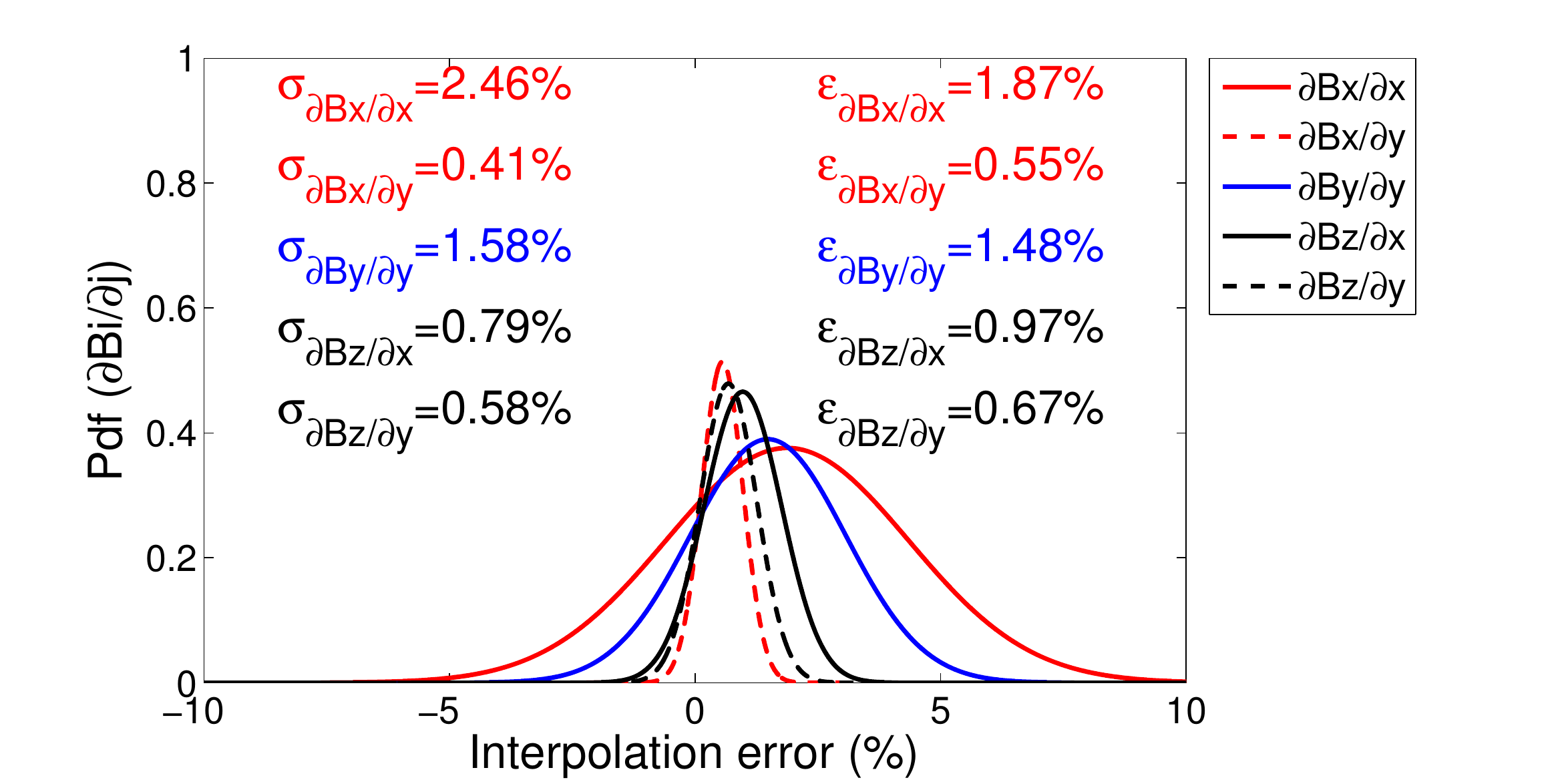}
\caption{Probability density function of the relative errors at the TM
  position  for $10^3$  random cases.  Five independent  terms in  the
  field   gradient  matrix   ($\partial  B_x/\partial   x$,  $\partial
  B_x/\partial y$,  $\partial B_y/\partial y$,  $\partial B_z/\partial
  x$  and   $\partial  B_z/\partial   y$)  are   considered.  Standard
  deviations  and  averaged  errors  relative to  the  modulus  (${\bf
  |\partial B/\partial x|}$ and  ${\bf  |\partial B/\partial y|}$) are
  shown.}
\label{fig:histogramGradB}
\end{figure}

Magnetic field gradients  also need to be estimated  from the readouts
of the  8 AMRs.  We do  this using  the multipole  expansion algorithm
because,   as   demonstrated   earlier,  this   interpolating   method
outperforms  the other  two methods  studied  here.  For  the sake  of
clarity,  only  the  errors  of the  gradient  interpolation  for  two
components ($\partial  B_x/\partial x$ and $\partial  B_z/\partial x$,
respectively)     along     the     spacecraft    are     shown     in
figure\,\ref{fig:errorylm8Grad}. In  this case, minimum errors  are
also
obtained in the center of the  TM, though unlike that obtained for the
case of the magnetic field, the error increases somewhat faster in the
region  outside   of  the  boundary   of  the  area   surrounding  the
magnetometers.  Additionally,  relative errors  around the TM  area are
slightly  larger  than  those  found for  the  reconstruction  of  the
magnetic  field, although  they  remain lower than $3\%$.
Figure\,\ref{fig:histogramGradB} shows the distribution of the
estimation errors and standard deviations for  five independent
components of the
gradient matrix $\Nabla{\bf B}$ at the position of the TM.  Inspection
of  this figure  reveals  that the  multipole expansion scheme  is
robust.  In particular,  when this  interpolant is  used we
obtain not only  accurate values of the  reconstructed magnetic field,
but also  of its  gradient, with  typical accuracies  of the  order of
$2\%$, and deviations below $2.5\%$ respectively.

\subsection{Other sources of error}

Absolute errors and drifts of  the magnetometers readings are relevant
to the interpolation quality, since the algorithm is entirely based on
the magnetometer outputs. Due  to the stringent stability requirements
for  eLISA, drifts  of the  measurements are  not critical.  Thus, the
analysis  is  focused  on  the   absolute  errors.   To  validate  the
robustness of the  system, the performance of  the multipole expansion
scheme is studied  for two common sources of  error.  Namely, possible
offsets in the magnetometer readings  and spatial uncertainty --- that
is,  deviations from  the  nominal  position of  the  sensor core.  We
analyze their eventual effects separately. Offsets in the magnetometer
or in the signal conditioning circuit can   be  measured   on-ground
and   considered  in   the analysis.  However,  unknown  magnetometer offsets
due  to  launch  stresses  can  lead to inaccurate field determination\,\cite{bib:offsetMag}. 
The precision of the position of the sensors may eventually be another source of error that
cannot be ignored {\em a priori}. The spatial uncertainty depends on the
size  of the  sensor head,  since smaller  heads result  in a  smaller
uncertainty of the precise location of the measurement.

The  offsets of  the magnetometers  can be  relevant depending  on the
measurement  technique.   In  particular, for  AMR  sensors,  flipping
signals applied to the sensor help to overcome the offset by reversing
the sensor magnetization and modulating the output signal\,\cite{bib:flipping}. The changes
in the direction of the sensor  magnetization lead to inversion of the
output characteristics  but not the  offset, which can be  canceled by
subtracting the measurements between  each flipping pulse.  Regarding the  spatial uncertainty, the
layout    of   the    thin   film    forming   the    AMR   Wheatstone
bridge\,\cite{bib:AMR} is deposited by a sputtering process, and has a
rough  area of  $0.9  \times 1.2\,{\rm  mm}^2$.  Therefore, a  spatial
uncertainty smaller than $1\,{\rm mm}$ is expected.

\begin{figure}[t!]
\begin{center}
\includegraphics[width=0.49\textwidth]{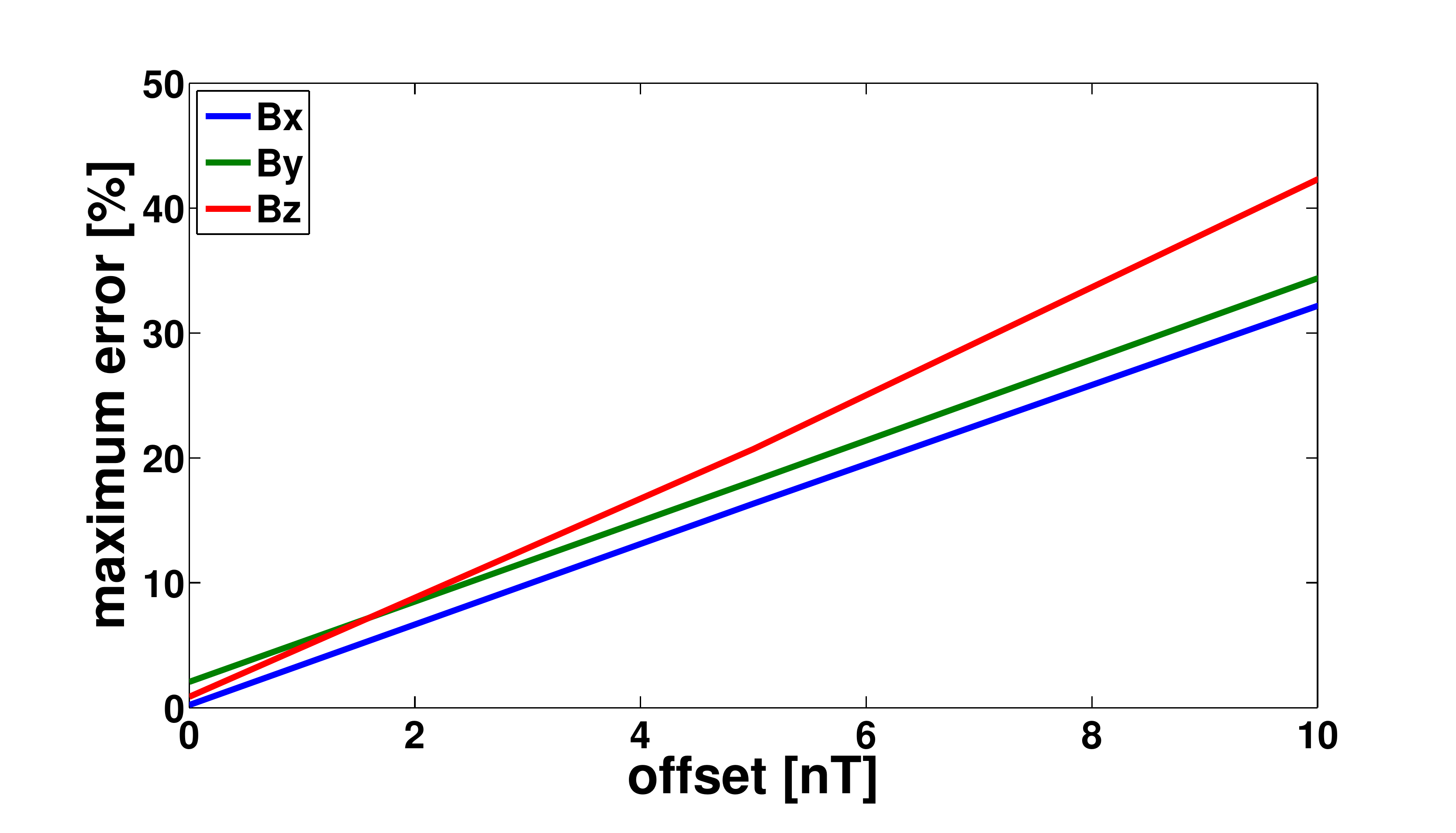}
\includegraphics[width=0.49\textwidth]{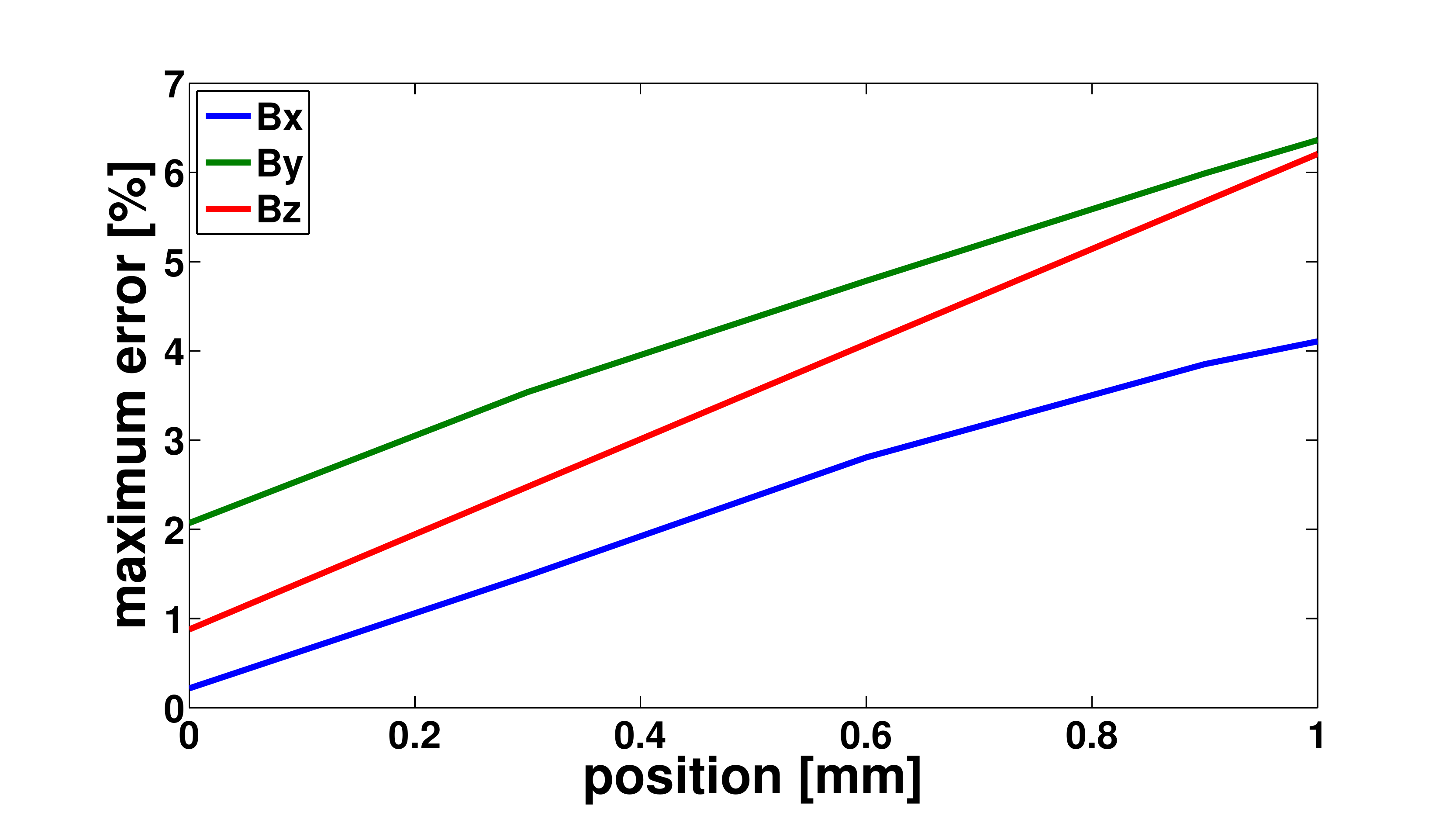}
\caption{Maximum estimation error of the  magnetic field as a function
  of  the  offset  (left)  and  spatial  uncertainty  (right)  of  the
  magnetometer.\label{ErrorMag}}
\end{center}
\end{figure}

The impact of these effects on the accuracy of the multipole expansion
algorithm is  simulated as follows.  First,  a $3 \times N$  matrix of
offsets is randomly generated according to a uniform distribution with
an interval of $[-B_{\rm Offset}, B_{\rm Offset}]$. Second, the offset
array is  added to the  $3 \times  N$ magnetic channels  readings, and
finally the magnetic field and  errors are estimated.  These steps are
sequentially  repeated  for  series  of  $10^3$  random  offsets  with
intervals of  the same length. A  similar procedure is done  to assess
the robustness of the interpolation to the uncertainty in the location
of the  sensor heads. The maximum  estimation errors as a  function of
the   offset   and  of   the   spatial   uncertainty  are   shown   in
figure\,\ref{ErrorMag}. As can  be observed, the offset  of the sensor
is more determinant than its  spatial resolution. Specifically, for an
unpredictable  non-measured  offset  of $10\,{\rm  nT}$,  the  maximum
estimation error is $\sim42\%$. These results reflect the relevance of
the  magnetic  sensing  technology.    Specifically,  we  stress  that
appropriate techniques to  cancel out the undetermined  offset and the
use  of tiny  sensors  with accurate  spatial  resolution are  totally
necessary.

      
\section{Conclusion}
\label{sec:conclusions}

 An AMR-based magnetic diagnostics subsystem for eLISA has been presented as an
alternative to the one using fluxgates in LISA Pathfinder. This new design
 leads to a reliable estimation of the  magnetic field and its gradient   
at the positions of the  test masses.  Actually, the  multipole expansion scheme
used in combination  with the proposed  8-sensor configuration
will represent a  reduction of the magnetic field  estimation error of
more  than two  orders of  magnitude when  compared to  the solution
implemented in LISA Pathfinder.  Besides,  we have shown that the
estimation errors computed for  different simulated magnetic scenarios
employing  the  multipole  expansion  interpolation  provides a robust
algorithm  that  does not  need  any  {\em a  priori} knowledge  of  the
magnetic  structure in  the spacecraft.   Also, in  addition to  these
significant advantages, the proposed system has the ability to deliver
correct  results  under  unpredictable  offsets  of  the  magnetometer
readings,  and  to overcome  reasonable  imprecisions  in the  spatial
location of the magnetometers.  All  in all, these improvements in the
accuracy of the magnetic field  reconstruction are achieved due to the
smaller  size and  lower magnetic  back-action of  the AMR  sensors,
which enable more sensors to be placed and for them to be located closer to the TMs.
This is a promising result that  proves that the use of AMRs combined
with  the  multipole expansion will provide a reliable estimate of
the magnetic characteristics  at the positions of the test masses of
eLISA.
      
\ack We are grateful to Airbus Defence and Space for providing the measured
magnetic moment values  for the LISA  Pathfinder units.  Support for  this work
came  from  grants  AYA2010-15709  and AYA2011-23102  of  the  Spanish
Ministry of Science  and Innovation (MICINN), ESP2013-47637-P of the Spanish Ministry of Economy and Competitiveness (MINECO), and 2009-SGR-935 (AGAUR). This  work was partially
supported by the European Union FEDER funds.

\end{document}